\documentclass[12pt]{article}
\pdfoutput=1
\usepackage{amssymb}
\usepackage{amsmath}
\usepackage{amsfonts}
\usepackage{graphicx}
\usepackage{mathrsfs}
\usepackage{comment}
\usepackage{cite}

\newcommand{\Slash}[1]{{\ooalign{\hfil#1\hfil\crcr\raise.167ex\hbox{/}}}}

\newcommand{\beq}{\begin{equation}}  \newcommand{\eeq}{\end{equation}}
\newcommand{\bef}{\begin{figure}}  \newcommand{\eef}{\end{figure}}
\newcommand{\bec}{\begin{center}}  \newcommand{\eec}{\end{center}}
\newcommand{\non}{\nonumber}  
\newcommand{\laq}[1]{\label{eq:#1}}  

\newcommand{\Eq}[1]{Eq.~(\ref{eq:#1})}
\newcommand{\Eqs}[1]{Eqs.~(\ref{eq:#1})}
\newcommand{\eq}[1]{(\ref{eq:#1})}

\newcommand{\ab}[1]{\left|{#1}\right|}
\newcommand{\vev}[1]{ \left\langle {#1} \right\rangle }

\newcommand{\U}[1]{{\rm U{#1} } }
\newcommand{\SU}[1]{{\rm SU{#1} } }
\newcommand{\SO}[1]{{\rm SO{#1}} }
\def\o{\over}
\def\a{\alpha}

\def\d{\delta}

\def\g{\gamma}

\def\l{\lambda}
\def\m{\mu}

\def\p{\psi}

\def\t{\tau}

\def\L{\Lambda}

\def\tl{\tilde}
\def\*{\dagger}

\def\({\left(}
\def\){\right)}

\def\diag{\mathop{\rm diag}\nolimits}

\def\O{\mathcal{O}}

\def\tr{\mathop{\rm tr}}

\newcommand{\OR}{~{\rm or}~}
\newcommand{\AND}{~{\rm and}~}

\newcommand{\GEV}{ {\rm ~GeV} }

\begin{document}
\begin{center}

\vspace{1.5cm}

{\Large\bf  Scale and quality of Peccei-Quinn symmetry and weak gravity conjectures}

\vspace{1.5cm}

{\bf Wen Yin} 

\vspace{12pt}

{\em  Department of Physics, Faculty of Science, The University of Tokyo, \\
Bunkyo-ku, Tokyo 113-0033, Japan,\\  Department of Physics, Korea Advanced Institute of Science and Technology, Daejeon 34141, Korea}

\abstract{
The promising solution to the strong CP problem by a Peccei-Quinn (PQ) symmetry may introduce quality and hierarchy problems, which are both relevant to Planck physics. In this paper, we study whether both problems 
can be explained by introducing a simple hidden gauge group that satisfies a weak gravity conjecture (WGC) or its variant. As a concrete example, we point out that a weakly-coupled hidden $\SU(N)$ gauge symmetry, which is broken down to $\SO(N)$, can do this job in the context of a Tower/sub-Lattice WGC. Cosmology is discussed.

}

\end{center}
\clearpage
\setcounter{footnote}{0}

\section{Introduction}
A global Peccei-Quinn (PQ) symmetry,  $\U(1)_{\rm PQ}$, is a leading candidate to solve a fine-tuning problem, the strong CP problem, of the standard model (SM) \cite{Peccei:1977hh,Peccei:1977ur}.  
Through the spontaneously breaking of $\U(1)_{\rm PQ}$, an axion, which is a pseudo-Nambu Goldstone boson (pNGB), arises~\cite{Weinberg:1977ma,Wilczek:1977pj,Kim:1979if,Shifman:1979if,Dine:1981rt,Zhitnitsky:1980tq}. Since the anomaly of $\U(1)_{\rm PQ}\text{-}\SU(3)_C^2$ is non-vanishing, 
 the axion gets a potential with a CP-conserving minimum due to the non-perturbative effect of the QCD, and thus at the vacuum the strong CP problem is solved.   
Because of the coherent oscillation in the early universe, the axion condensate can contribute to the matter density and hence can explain the dark matter~\cite{Abbott:1982af,Dine:1982ah, Preskill:1982cy}.
(See e.g. Refs.~\cite{Kim:2008hd,Wantz:2009it,Ringwald:2012hr,Kawasaki:2013ae,Marsh:2015xka, Graham:2015ouw,  Irastorza:2018dyq, DiLuzio:2020wdo} for reviews.)

The solution, however, suffers from hierarchy and quality problems. The first problem is due to that the PQ scale, or the decay constant of the QCD axion, $f_a$, is constrained to be within the so-called classical axion window: 
\beq
\laq{window}
10^{8}\GEV \lesssim f_a\lesssim 10^{12}\GEV,
\eeq
which is  smaller than the reduced Planck scale, $M_{\rm pl}=2.4\times 10^{18}\GEV$.
The lower bound comes from the duration of the neutrino burst in the SN1987a~\cite{Chang:2018rso} (See also Refs.~\cite{Mayle:1987as, Raffelt:1987yt}).
The upper bound comes from the axion abundance constraint.
One simple way to address this hierarchy is to open the window. This is possible if the Hubble parameter during the inflation, which lasts long enough, is lower than the QCD scale~\cite{Graham:2018jyp, Guth:2018hsa}.\footnote{This low scale inflation can also alleviate the moduli problem at the same time~\cite{Ho:2019ayl}. See also related topics~\cite{Tenkanen:2019xzn, Co:2018phi, Okada:2019yne, AlonsoAlvarez:2019cgw, Marsh:2019bjr, Matsui:2020wfx, Nakagawa:2020eeg}. It is also possible to introduce other degrees of freedom to, e.g., dilute or transfer the axion abundance~\cite{Dine:1982ah, Steinhardt:1983ia, Kitajima:2014xla, Kawasaki:2015lpf, Kitajima:2017peg, Agrawal:2017eqm}.} 
Another way is to extend the PQ sector to make the scale of the classical axion window natural, e.g. the axion is composite~\cite{Kim:1984pt,Choi:1985cb}, or with supersymmetry. Our proposal will belong to the latter. 

The quality problem, on the other hand,  is somewhat related to quantum gravity. It is believed that any global symmetry should be explicitly broken by Planck-scale physics (See e.g. Refs.~\cite{Misner:1957mt, Barr:1992qq, Banks:2010zn}).
Thus the PQ symmetry should be explicitly broken. 
It was pointed out that a PQ symmetry with good enough quality can be obtained as an accidental symmetry of discrete gauge symmetries~\cite{Chun:1992bn,BasteroGil:1997vn,Babu:2002ic,Dias:2002hz, Harigaya:2013vja}, abelian gauge symmetries~\cite{Fukuda:2017ylt,Duerr:2017amf,Bonnefoy:2018ibr}, and non-abelian gauge symmetries~\cite{Randall:1992ut,DiLuzio:2017tjx, Lillard:2018fdt, Lee:2018yak}.  
A relevant criterion for quantum gravity is weak-gravity conjecture (WGC)~\cite{ArkaniHamed:2006dz}, which suggests that gravity is the weakest long-range force.  
From this conjecture, given the charged particle spectrum, we cannot take the coupling of an unbroken gauge symmetry to be arbitrarily small, especially zero to get a continuous global symmetry. In other words, given the gauge coupling, the particle spectrum is constrained.
In fact, by introducing an unbroken $\U(1)_{\rm B-L}$ symmetry which is weakly coupled, the hierarchy between the electroweak and Planck scales was discussed within the WGC~\cite{Cheung:2014vva}. 
It may be also important to discuss the scale of the PQ symmetry in the context of the WGC via the gauge symmetry introduced for the quality problem.

In this paper, we find that by introducing a $\U(1)'$ gauge symmetry in a KSVZ axion model~\cite{Kim:1979if,
Shifman:1979if}, 
 the hierarchy between the scale of axion window and the Planck scale can be explained within the context of the WGC, which restricts an unbroken abelian gauge symmetry. 
 However it cannot solve the quality problem because if the PQ symmetry appears as the accidental symmetry of $\U(1)'$, $\U(1)'$ must be broken in order to have the spontaneous PQ symmetry breaking.
From this finding, we study a fundamental axion model with a hidden $\SU(N)$ gauge symmetry, which is incompletely broken down via the PQ symmetry breaking, and solves the quality problem. 
In this case, a Tower/Sub-Lattice WGC (sLWGC)~\cite{Heidenreich:2015nta,
Heidenreich:2016aqi,
Montero:2016tif, Andriolo:2018lvp} can set a cutoff to the energy scales of the field theory~\cite{Heidenreich:2017sim}. 
If the cutoff is around the axion window, the PQ scale cannot be higher than that and, as a result,  
both the quality and hierarchy problems are solved. We also point out that within this low-cutoff theory, one may have a consistent cosmology. 

This paper is organized as follows. In the next section we review the hierarchy problem for the PQ scale, and discuss the possible solution by introducing an unbroken abelian gauge symmetry with mild-version of WGC. 
In the Sec.\,\ref{sec:3}, we study the non-abelian gauge theory with tower/sLWGC and show that both the quality and hierarchy problems can be solved. The cosmology of the scenario is also discussed. 
The last section is devoted to conclusions and discussion.

\section{Scale of fundamental axion and WGC}

\subsection{Hierarchy problem of the PQ scale}
Let us consider a KSVZ model with the following particle contents 
\beq
\laq{fax}
q: (1, r_{\rm SM}),~ \bar{q}: (0 , \bar{r}_{\rm SM}), ~H_{\rm PQ}: (-1, 1)\eeq 
Here $q\AND \bar{q}$ are exotic PQ quarks, and $H_{\rm PQ}$ is a PQ Higgs field, under the representation of $(\U(1)_{\rm PQ}, {\cal G}_{\rm SM}).$
$r_{\rm SM}$ denotes the representation of the SM gauge group ${\cal G}_{\rm SM}=(\U(1)_Y,\SU(2)_L,\SU(3)_C)$, and we take $r_{\rm SM}=(Y, 1_L, 3_C)$ here and hereafter. 
Then it is allowed to write down the Yukawa coupling of 
\beq
{\cal L} \supset  y_q \bar{q} H_{\rm PQ} q. 
\eeq
The potential of the $H_{\rm PQ}$ is given as 
\beq
V_{\rm PQ}=-m_{\rm PQ}^2 \ab{H_{\rm PQ}}^2+{\l\over 2} \ab{H_{\rm PQ}}^4.
\eeq
$m_{\rm PQ}^2(>0)$ and $\l$ are the mass parameter and quartic coupling, respectively. 
Thus one readily gets that the $H_{\rm PQ}$ obtains a non-vanishing vacuum expectation value (VEV)
as 
\beq
\laq{vev}
v_{\rm PQ}\equiv \vev{H_{\rm PQ}}={m_{\rm PQ}\over \sqrt{\l}}.
\eeq
The quarks get mass of
\beq
\laq{Mq}
M_q=y_q v_{\rm PQ}.
\eeq
Then the PQ symmetry is spontaneously broken and 
a (pseudo) NGB, $a$, appears which couples to  $q \AND \bar{q}$. 
Since the anomaly of $\U(1)_{\rm PQ}\text{-}{\cal G}^2_{\rm SM}$ is non-vanishing, 
by integrating out the heavy quarks one obtains 
\beq
\laq{ancp}
{\cal L}^{\rm eff}\supset {1 \over 16\pi^2}{a\over \sqrt{2} v_{\rm PQ}}\( {3Y^2} g_Y^2 F_Y\tl{F}_Y+
   g_3^2 \tr[F_C\tl{F}_C]\). 
\eeq
Thus $a$ is the QCD axion and 
\beq
\laq{const}
f_a=\sqrt{2}
 v_{\rm PQ}\eeq
which should satisfy the classical axion window \eq{window}.

Next, let us introduce a hierarchy problem of this model,
which can be regarded as a radiative instability problem. 
The radiative correction to the mass parameter of the PQ field is 
\beq
\d m_{\rm PQ}^2 = \O({\max{[\l, y^2_q ]}}){\L^2_{\rm c.o.}\over 16\pi^2}
\eeq
 where  $\L_{\rm c.o.}$ is the cutoff scale of this model.  We will take into account the quantum gravity effects, which we have omitted here, later.
By assuming the cutoff scale $\L_{\rm c.o}\sim M_{\rm pl}$, the radiative correction is $10^{17}\GEV$ for $\O(1)$ couplings. 
This implies that $m^2_{\rm PQ}$ in the conventional axion window \eq{const} 
requires fine-tuning of 
$
m_{\rm PQ}^2/\d m_{\rm PQ}^2\lesssim 10^{-10}
$ between the bare mass squared and $\d m_{\rm PQ}^2$.
Notice that by simply taking the coupling small the tuning cannot be relaxed. 
Since $v_{\rm PQ}^2= \l^{-1}m_{\rm PQ}^2,$ a small $\l$ implies an even smaller $m_{\rm PQ}^2,$ i.e. the ratio $\d m_{\rm PQ}^2/m_{\rm PQ}^2$  remains.

\subsection{WGC and PQ scale: case of $\U(1)'$ }

The WGC states that gravity is the weakest long-range force. 
More precisely, it says that in the effective theory of $\U(1)'$ gauge symmetry consistent with the quantum gravity, there is at least a charged particle with mass $m$ and charge $q'$ satisfying~\cite{ArkaniHamed:2006dz}
\beq
\laq{WGC}
m\lesssim q' g' M_{\rm pl}~ ({\rm WGC})
\eeq
where $g'$ is the coupling of the $\U(1)'$ at the scale of $m$. In this section, we take $q'=1$ for simplicity. 
If the WGC is violated, charged black holes would become stable which is unnatural. 
There is not yet any counterexample found in string theory to (the mild version of) the WGC. 
Moreover, there are several proof of the WGC under certain assumptions~\cite{Hod:2017uqc,
Fisher:2017dbc,
Montero:2018fns,
Cheung:2018cwt,Hamada:2018dde}. 

Now let us follow \cite{Cheung:2014vva} to explain the hierarchy by using \Eq{WGC}. 
Let us introduce an unbroken $\U(1)'$ gauge symmetry under which 
\beq
q : 1~~ \bar{q}: -1.
\eeq
The other fields (including the SM particles) are all supposed to be charge-less for simplicity.\footnote{One notices that the anomalies of $\U(1)_{\rm PQ}\text{-}\U(1)_Y\text{-}\U(1)'$ and $\U(1)_{\rm PQ}\text{-}\U(1)'^2$ are non-vanishing if $Y$ is non-vanishing.
This, and possible kinetic mixing to a photon, may lead to the axion couplings to the hidden photon field strength, $F'$, as
$
{\cal L}^{\rm eff}\supset { 3Y g_Y g' \over {16\pi^2}v_{\rm PQ}} a F_Y\tl{F'} +{  3 g'^2 \over {16\pi^2}v_{\rm PQ}} a F'\tl{F'}\equiv {g_{a\g\g'}\over 4} a F_Y\tl{F'}+{  g_{a\g'\g'}\over 4} a F'\tl{F'}
$
which is studied in e.g. Refs.\,~\cite{Jaeckel:2014qea, Kaneta:2016wvf, Choi:2016kke, Daido:2018dmu}. } 
For later convenience, let us rewrite the gauge coupling 
\beq
\laq{wc}
g' \equiv {y_q} {\tl{f} \over  \sqrt{2}  M_{\rm pl} } \simeq 3\times 10^{-7}  y_q \({\tl{f}\over 10^{12}\GEV}\),
\eeq
by the dimensionful parameter $\tl{f}$. 

From \Eq{WGC}, 
it turns out that 
\beq
M_q\leq   y_q {\tl{f}\over \sqrt{2}}. 
\eeq
Consequently, from \Eq{Mq}, one obtains that
\beq
v_{\rm PQ} \lesssim {{\tl{f}}\over \sqrt{2}}
\eeq
Notice that this bound is from the consistency condition with quantum gravity if the WGC is correct, and thus the na\"{i}ve radiative instability discussion neglecting quantum gravity does not apply. 
The mildest tuned mass parameter of PQ Higgs satisfies 
\beq
m_{\rm PQ}\simeq \sqrt{\l \over 2} {\tl{f}},
\eeq
i.e. the WGC bound is saturated.\footnote{Strictly speaking, the total gauge group including the ${\cal G}_{\rm SM}$ is a product group, and one should apply the convex hull condition~\cite{Cheung:2014vva}.
We have checked that this does not change much our conclusion as an order of estimate.}
As a result, if $\tl{f}$ is within or slightly above the axion window,  $f_a$ can be naturally within \eq{window}.

In the explanation of the hierarchy between 
the $v_{\rm PQ}$ and $M_{\rm pl}$, we have introduced a small parameter $g'$. 
Although the small parameter is technical natural, the mildest tuned parameter set is 
\beq
g'\sim 10^{-6}~ ({\rm for}~ \tl{f} \simeq 10^{12}\GEV), ~y_q=\O(1).
\eeq

One may wonder if $\U(1)'$ with a proper charge assignment can lead to an accidental PQ symmetry, 
and solve the quality problem. However, it is difficult.  
If the PQ symmetry were an accidental symmetry relevant to $\U(1)'$, 
a PQ Higgs field, which breaks the PQ symmetry, should be also charged under $\U(1)'$. 
Therefore $\U(1)'$ symmetry must be broken, which means that the WGC cannot apply. 
To sum up, in the context of the mild version of the WGC, an unbroken $\U(1)'$ symmetry can solve the hierarchy problem of the PQ symmetry but cannot solve the quality problem.\footnote{One option is to introduce another gauge symmetry to solve the quality problem, namely the quality and hierarchy problems are solved by different gauge symmetries.  }

\section{PQ scale and quality with non-abelian hidden gauge symmetry }

\label{sec:3}

To have an unbroken gauge symmetry relevant to the quality of the accidental PQ symmetry, we will introduce a non-abelian gauge symmetry instead of $\U(1)'$.
This gauge symmetry is spontaneously broken down to an unbroken gauge symmetry via the PQ symmetry breaking.  
In this case, the WGC can apply to the remnant unbroken gauge symmetry.

\subsection{A hidden $\SU(N)$ gauge model for precise PQ symmetry}
To be concrete, let us assume that the exotic quarks are charged under a hidden $\SU(N)$ gauge group. 
The charge assignments of $(\SU(N),{\cal G}_{\rm SM})$ are given as
\beq
\laq{fax}
q: (\bar{N}, r_{\rm SM}),~ \bar{q}: (\bar{N}, \bar{r}_{\rm SM}),~ \p_a: (N,1 )
\eeq
where $N$ is the fundamental representation of $\SU(N),$ and $\p_a$ are needed to cancel the gauge anomaly of $\SU(N)^3$ 
with $a= 1\cdots 2\dim{[r_{\rm SM}]}.$

To give masses to the exotic quarks let us introduce a Higgs field who is a symmetric tensor of the second rank,
\beq
\laq{H2}
H_{\rm PQ}\equiv H^{\{ij\}}_{\rm PQ}: \({N^2+N\o 2}, 1\).
\eeq
We have explicitly written the symmetric indices, $i,j=1\cdots N$ of $\SU(N)$. 
(We consider this representation because of simplicity, and  because that $N$ required for the PQ quality, as discussed below, is smallest. From the discussion in Ref.\,\cite{Lee:2018yak}, it is easy to consider other 
representations with good PQ quality. Another simple possibility is discussed in Appendix \ref{sec:ap}, in which however, $\SU(N)$ is completely broken.)
The renormalizable Yukawa terms are given by 
\beq
{\cal L} \supset  y_q \bar{q} H_{\rm PQ} q+ y_\p^{ab} \p_aH^*_{\rm PQ}\p_b
\eeq
where $y_q, \AND y_\p$ are the Yukawa couplings. 

In fact, it was pointed out in Ref.~\cite{Lee:2018yak} that a large $N$ $\SU(N)$ gauge theory can generically lead to an accidental $\U(1)_{ B_H}$ global symmetry (hidden baryon number symmetry) originating from the 
$N$-ality due to the group structure. 
The $\U(1)_{B_H}$ charge assignment is automatically obtained as 
\beq q,\bar{q} : -1, ~\p_a: 1,~ H_{\rm PQ}: 2,\eeq 
by counting the number of the 
indices with the sign corresponding to the complex representation. 
One can check that in the Yukawa term and the following Higgs potential that $\U(1)_{B_H}$ manifests with good quality. 
The leading operator that breaks $U(1)_{ B_H}$ has a dimension $N$ as
\beq
\laq{HBB}
{\cal L}\supset c_N {\det{[H_{\rm PQ}]} \over M_{\rm pl}^{N-4}}.
\eeq
Here $c_N$ is a constant and this term may be generated through a quantum gravity effect~(e.g. Refs.~\cite{Abbott:1989jw, Coleman:1989zu, Banks:2010zn}). 
One notices that 
\beq
\U(1)_{B_H}\text{-}{\cal G}_{\rm SM}^2\supset \U(1)_{B_H}\text{-}\SU(3)_c^2
\eeq
is anomalous. 
It turns out that the $\U(1)_{B_H}$ is the PQ symmetry
\beq
\U(1)_{\rm PQ}\equiv \U(1)_{B_H}.
\eeq
The PQ symmetry can be precise enough against Planck-scale suppressed terms and solve the strong CP problem if \beq N\gtrsim 9\eeq with $c_N=\O(1), f_a=10^8\GEV$~\cite{Barr:1992qq}. 
Consequently, the notorious quality problem of the PQ symmetry can be solved.

It may be non-trivial whether the $\U(1)_{\rm PQ}$ can be spontaneously broken down, although we have implicitly assumed.
The potential of the $H_{\rm PQ}$ is obtained as 
\begin{align}
V={\l_1\over 4} \tr[&H^\*_{\rm PQ}H_{\rm PQ}]^2 +{\l_2\over 4} \tr[(H^\*_{\rm PQ}H_{\rm PQ})^2]\non\\
&-{m_{\rm PQ}^2\over 2} \tr[H^\*_{\rm PQ}H_{\rm PQ}].
\end{align}
Here $\l_1,\l_2$ are quartic couplings.
At the minimum of the potential, one obtains the non-vanishing VEV
\beq
\vev{H_{\rm PQ}}= v_{\rm PQ}\diag{[1,\cdots 1]} ~~{\rm if~\l_2>0}
\eeq
[When $\lambda_2\leq 0$, $\vev{H_{\rm PQ}}\propto\diag{[1,0\cdots 0]}$, which we do not consider throughout the paper.]
Here, 
\beq v_{\rm PQ}^2={m_{\rm PQ}^2 \over N \l_1+\l_2}.\eeq
Thus, 
\beq
\laq{sym2}
\SU(N) \times \U(1)_{\rm PQ} \rightarrow \SO(N)\times Z_2,
\eeq
where $Z_2$ is the remnant of the $\U(1)_{\rm PQ}$. Thus, $H_{\rm PQ}$ not only breaks $\SU(N)$ but also $\U(1)_{\rm PQ}$ incompletely. 
A QCD axion appears with the coupling of
\beq
\laq{ancp}
{\sqrt{N} \over 16\pi^2}{a\over v_{\rm PQ}}\( 3Y^2 g_Y^2 F_Y\tl{F}_Y+
   g_3^2 \tr[F_C\tl{F}_C]\) . 
\eeq
It turns out that 
\beq
\laq{rel}
f_a={1\over \sqrt{N}} v_{\rm PQ}.
\eeq

\subsection{WGC and PQ scale: case of $\SO(N)$ }
Since there is an unbroken gauge symmetry $\SO(N),$ the WGC can apply. However the mild-version of the WGC for any $\U(1)$ in the subgroup of $\SO(N)$ is
 satisfied due to the charged massless gauge bosons of $\SO(N)$ and the charged massive gauge bosons eating the massless NGBs in $H_{\rm PQ}$.
On the other hand, it is considered that the WGC should be somewhat sharpened. One reason is that the mild-version is not invariant under the dimensional reduction.

The Tower/sub-lattice WGC (sLWGC)~\cite{Heidenreich:2015nta,
Heidenreich:2016aqi,
Montero:2016tif, Andriolo:2018lvp}, belonging to the stronger variants of the WGC motivated by the invariance under dimensional reduction, states:
an infinite tower of particles/resonances of different charges satisfying \eq{WGC}  exists. 
This conjecture also clears various theoretical tests.
Since a large number of particles exist, they come into the loop of gravity and makes the gravity strongly coupled at a scale $\L_{\rm QG}$. $\L_{\rm QG}$ satisfies~\cite{ArkaniHamed:2005yv} \beq\laq{GCUT} M_{\rm pl}\gtrsim \sqrt{N_{\rm states}} \L_{\rm QG}, \eeq, where $N_{\rm states}$ is the number of states below the 
$\L_{\rm QG}$. $\L_{\rm QG}$ can be seen as the cutoff scale for the quantum field theory. 
If $N$ is so large that the number of states in the tower increases fast enough~\cite{Heidenreich:2017sim}, 
\beq
\laq{sLWGC}
\L_{\rm QG}\lesssim g_N M_{\rm pl} 
\eeq
where $g_N$ is the coupling of the large $N$ gauge theory at the scale $g_N M_{\rm pl}$. 
In the sLWGC, $\log{N_{\rm states}}\sim {N^2}\log{(\Lambda_{\rm QG}/g_N M_{\rm pl})}$ for large $N$ and one can get \eq{sLWGC} from \eq{GCUT}~\cite{Heidenreich:2017sim}.

Now let us discuss the hierarchy between the $m_{\rm PQ}$ and $M_{\rm pl}.$
Since there is an unbroken ${\rm SO}(N)$ gauge symmetry, following \eq{sLWGC} one gets the cutoff of the quantum field theory of
\beq \L_{\rm QG} \lesssim 10^{13}\GEV \({g_N \o 10^{-5}}\). \eeq
By identifying 
\beq
\L_{\rm c.o.}=\L_{\rm QG},
\eeq
the radiative correction is then given by 
\beq
\d m_{\rm PQ}^2 = \O({\max[|\l_1| N^2, |\l_2|, \ab{y_q}^2, \ab{y_\p}^2, g^2_N] }){\L^2_{\rm QG}\over 16\pi^2}.
\eeq
There is no fine-tuning between the radiative correction and the bare mass if
\beq
\laq{NAB2}
\O({\L_{\rm QG} \over \sqrt{16\pi^2}}) \lesssim v_{\rm PQ} \lesssim \L_{\rm QG}.
\eeq
When the lower bound is satisfied, \beq f_a \sim 2\times 10^{11}\GEV {\sqrt{13\over N}}\( {\L_{\rm QG}\o 10^{13}\GEV}\). \eeq
Consequently, both the quality and hierarchy problems of the PQ symmetry are solved in this model in the context of the Tower/sLWGC.

\subsection{Cosmology}

The cosmology of models solving the quality problem is usually troublesome. 
For instance, in our model $Z_2$ symmetry in \eq{sym2} stabilizes $q,\bar{q}, \AND \p_a$ which could cause cosmological problem\footnote{This problem may be solved if we consider $q: (\bar{N}(\bar{N}+1)/2, r_{\rm SM}),~ \bar{q}: (\bar{N}(\bar{N}+1)/2, \bar{r}_{\rm SM}),~ \p_a: ({N}({N}+1)/2,1 )$ instead. They have charges $2$ and $-2$ under $\U(1)_{\rm B_H}$, and thus neutral under the $Z_2$ symmetry. 
In this case, dimension 4 or 5 terms involving SM particles are allowed if $Y=\pm 1/3 \OR \pm 2/3.$
}, although some of the fermion could be the dark matter. Also, the PQ symmetry breaking, if happens during the thermal history, may generate domain walls. If the PQ symmetry is not restored during inflation and the axion is the dominant dark matter, there can be an isocurvature problem. 
These problems are all solved if the inflation scale is small enough. However, again, a hierarchy between the inflation and the Planck scales is introduced. 

In our scenario, the inflation scale must be smaller than $\L_{\rm QG}$,  
i.e. the Hubble parameter during inflation, $H_{\rm inf}$, satisfies 
\beq
H_{\rm inf}< 2\times 10^7\GEV \(\frac{\L_{\rm QG}}{10^{13}\GEV}\)^2.
\eeq
Thus the above problems can be solved with small inflation scales with the hierarchy explained in the context of the Tower/sLWGC and the marginally small $g_N.$
A direct prediction of the scenario is the suppressed tensor/scalar ratio, 
\beq
 r \approx 1.6 \times 10^{-15} \left( {H_{\rm inf }
    \over 10^7{\rm\, GeV}}\right)^2 .
\eeq
If we maximize the inflation scale, and if the QCD axion is dominant dark matter, the induced isocurvature perturbation is close to the current bound~\cite{Akrami:2018odb}. Thus it may be searched for in the near future. 

Interestingly, the cutoff scale $\L_{\rm QG}\sim 10^{13}\GEV$ is close to the seesaw scale. If the right-handed neutrino masses are around the cutoff scale, 
and the neutrino Yukawa couplings are not too small, one explains the active neutrino masses with correct scales via the seesaw mechanism~\cite{Yanagida:1979as,Glashow:1979nm, 
GellMann:1980vs,Minkowski:1977sc,Mohapatra:1979ia}. Thermal leptogenesis is possible~\cite{Fukugita:1986hr}.

Now let us come back again to the quality problem, which we have assumed that the PQ breaking terms appear from the Planck-scale-suppressed terms. 
On the other hand, the higher dimensional operator \eq{HBB} may be generated at the scale of $\L_{\rm QG},$ 
$
c_N\sim \({M_{\rm pl}\over  \L_{\rm QG}}\)^{N-4},
$
although it is model dependent (See c.f. Refs.~\cite{ArkaniHamed:1998sj, ArkaniHamed:1999dc}). With $v_{\rm PQ}/\L_{\rm QG}\sim \sqrt{1/16\pi^2}$, the quality can be good enough if 
$ \laq{qua} N\gtrsim 50.$
If the higher dimensional terms of the SM particles are generated at the scales of $\L_{\rm QG}$, 
the axion dark matter with $f_a\simeq v_{\rm PQ}/\sqrt{N}\sim10^{12}\GEV$ implies a cutoff of
$\L_{\rm QG} \sim 10^{13-14}\GEV.$
In this scenario, the neutrino masses can be generated by the higher dimensional operators of 
\beq
\laq{HD}
{c^{ab}\over  \L_{\rm QG} } {H_{\rm SM}L_a H_{\rm SM}L_b  }
\eeq
correctly for the dimensionless coefficient $c^{ab}\sim\O(1).$ Here $H_{\rm SM} \AND L_a$ are the SM Higgs and lepton doublet fields with the flavor indices $a,b=e,\m,\t.$ 
In this case, the baryon number violating operators may also exist although it again depends on the detail of the UV model (See e.g. Ref.\,\cite{Dienes:1998vg, ArkaniHamed:1998sj, ArkaniHamed:1999dc}). 
To avoid the proton decays, one can introduce a $Z_2$ symmetry under which SM leptons are odd but the baryons are even or vise versa. 
In this case, the baryon asymmetry may be correctly obtained from neutrino oscillation with the higher dimensional term \eq{HD}~\cite{Hamada:2018epb, Eijima:2019hey}
and also the quark oscillation could be important via the baryon-number-violating but baryon-parity-conserving operator~\cite{Asaka:2019ocw}.

Lastly, let us mention the confinement of $\SO(N)$. The unbroken ${\rm SO}(N)$ becomes non-perturbative at low energy scales. 
However, $g_N$ is tiny, which means that the ``QCD" scale 
$\Lambda_{{\rm SO}(N)} = e^{-{8\pi^2/b g_N^2}} f_a$ with $b=11(N-2)/3,$  is extremely low. 
For $N<10^9, g_N<10^{-5}, f_a\sim 10^{12}\GEV$, the size of the instanton 
$1/\L_{{\rm SO}(N)}$ is much larger than  $1/H_0$, where $H_0$ is the current Hubble parameter,  and thus we can neglect the non-perturbative effect in phenomenology.

\section{Conclusions and Discussion}

An anomalous PQ symmetry solves the strong CP problem, which is a fine-tuning problem. 
The solution, however, suffers from other fine-tuning problems, the quality and hierarchy problems. Moreover, to solves the potentially existing domain wall, stable PQ fermions, and isocurvature problems, a low inflation scale may be needed, 
 which may introduce another hierarchy problem.

In this paper, we have studied whether the hierarchy and the quality problems for the PQ symmetry can be both explained by introducing a simple hidden gauge group which satisfies the WGC or its variant.  
By introducing an unbroken $\U(1)'$ gauge symmetry under which some PQ fermions are charged, the mild version of the WGC can constrain the PQ scale to be below $10^{12}\GEV$ for the coupling $g' \lesssim 10^{-6}$. However, the quality problem cannot be solved. A non-abelian hidden gauge symmetry which is partially broken down via the spontaneous PQ breaking can solve the quality problem. 
In this case, according to a stronger version of the WGC, the Tower/sLWGC, the cutoff scale of the field theory is reduced to be around the QCD axion window if the coupling satisfies $g_N\lesssim 10^{-5}$. As a result the small PQ  and inflation scales can be simultaneously explained. Interestingly, the seesaw scale is also around the cutoff scale.

A $\L_{\rm QG}<10^{15-16}\GEV$ may not be compatible with the ordinary grand unified theory (cf. Ref.\,\cite{Dienes:1998vg}), which explains the quantized charges of the SM. 
However the charge quantization can be achieved if supersymmetry restores at the scale around $\L_{\rm QG}$ and a slepton is the SM Higgs field~\cite{Yin:2018qcs}, given the three generations. 

Although we have focused on the QCD axion, in general a light pNGB with a small decay constant, has both the problems of quality and hierarchy.  
Our mechanism can also apply to a general pNGB or ALP.\footnote{Such ALP may be important in various contexts e.g. ALP inflation~\cite{Daido:2017wwb,Daido:2017tbr, Takahashi:2019qmh}, and the explanation of the XENON1T excess~\cite{Aprile:2020tmw} \cite{ Takahashi:2020bpq,  Bloch:2020uzh, Li:2020naa,Athron:2020maw, Han:2020dwo, Takahashi:2020uio}. 
}
Then the tiny mass of the ALP can be generated via the higher dimensional terms or the non-perturbative effect of $\SO(N)$ dynamics if $\U(1)_{B_H}\text{-}\SO(N)^2$ is anomalous\footnote{For instance, we may change the fermion contents, \eq{fax} to $\p_\a: (N,1), \tl{\p}: (\bar{N}(\bar{N}+1), 1)$
where $\a=1\cdots N+4,$ we find $\U(1)_{B_H}\text{-}\SO(N)^2$ anomaly is non-vanishing. Here, $\tl{\p}$ has $\U(1)_{B_H}$ charge $2$ accidentally.
} and $g_N^2 N$ is large enough. The non-perturbative effect, if sizable, requires an extremely large $N$ for a small $\L_{\rm QG}.$
\\

{\it Note added:}
While preparing this paper, we found Ref.~\cite{Ardu:2020qmo} where the model with $\SU(N)$ hidden gauge symmetry broken down to $\SO(N)$ was discussed in the context of quality problem. 
In this paper, we mainly focus on the small $g_N$ regime, and find that due to the unbroken $\SO(N)$ the spectra may be restricted by weak gravity conjectures. 
We showed that the much lower PQ scale than $M_{\rm pl}$ and safe cosmology can be obtained due to the resulting low cutoff scale. 
These are not discussed in~\cite{Ardu:2020qmo}.

\section*{Acknowledgments}
The author thanks Prof. Hye-Sung Lee for discussion and reading the manuscript. 
 This work was supported by JSPS KAKENHI Grant No. 16H06490 and by National Research Foundation Strategic Research Program (NRF-2017R1E1A1A01072736).

\appendix
\section{Alternative models with sLWGC}
\label{sec:ap}
It was discussed that the cutoff $\L_{\rm QG}$ may be set even with the non-abelian gauge field completely broken by higgsing~\cite{Heidenreich:2017sim, Reece:2018zvv}. In this case, we can consider instead of \Eqs{fax} and \eq{H2}
\beq
\laq{fax2}
q: (\bar{N}, r_{\rm SM}),~ \bar{q}_l: (1, \bar{r}_{\rm SM}), ~ \p_a : 
(N, 1).
\eeq 
and 
\beq
H_{\rm PQ}^l :(N,1),
\eeq
where $l=1\cdots N$ and $a=1\cdots \dim[r_{\rm SM}]$. 
The Yukawa interaction is given by 
\beq
{\cal L} \supset  (y_q)^{m}_l \bar{q}_m H^l_{\rm PQ} q. 
\eeq
Again we get accidental $\U(1)_{\rm PQ}$ where 
$q: -1, ~ \p_a:1, ~H_{\rm PQ}^i: 1,~ \bar{q_l}: 0.$
Supposing that we have a potential at the minimum 
$\vev{(H_{\rm PQ}^l)_i}=v^i \delta^{l}_j\neq 0,$
we obtain 
$
\SU(N)\times \U(1)_{\rm PQ}
$
completely broken, and an axion appears. The axion couples to the gluon with the decay constant given by 
\beq
f_a= \sqrt{2\sum_{i}{\({ 1 \over v_i \sum_{j}{1/v_j^2}}\)^2}}.
\eeq 
The hierarchy are quality problems are similarly solved as in the main part. 

If a spontaneous broken $\U(1)'$ symmetry is also restricted by the sLWGC, the cutoff is set by $\L_{\rm QG}\lesssim (g')^{1/3}M_{\rm pl}$~\cite{Heidenreich:2017sim}.
Then both the quality and scale of the PQ symmetry can be explained with $g'\sim 10^{-15}$ and with certain charge assignment~\cite{Fukuda:2017ylt,Duerr:2017amf,Bonnefoy:2018ibr}.

\end{document}